\documentclass[aps,prl,twocolumn,superscriptaddress]{revtex4}

\usepackage{mathrsfs,amssymb,graphics,subfigure,threeparttable}
\usepackage{grap hicx}
\usepackage{amssymb}
\usepackage{enumerate}
\usepackage{amsmath}
\usepackage{fancyhdr}
\usepackage{bm}
\usepackage[squaren]{SIunits}
\usepackage{hyperref}
\usepackage{ulem}
\usepackage{textcomp}
\usepackage{color}
\renewcommand{\Re}{\operatorname{Re}}

\begin{document}


\title{Efficient generation of intense spatial and spatiotemporal vortex harmonics using plasma mirrors}

\author{Yipeng Wu}
\email{wuyipeng@ucla.edu}
\affiliation{University of California, Los Angeles, California 90095, USA}
\author{Zan Nie}
\affiliation{University of California, Los Angeles, California 90095, USA}
\author{Fei Li}
\affiliation{University of California, Los Angeles, California 90095, USA}
\author{Chaojie Zhang}
\affiliation{University of California, Los Angeles, California 90095, USA}
\author{Ken A Marsh}
\affiliation{University of California, Los Angeles, California 90095, USA}
\author{Warren B. Mori}
\affiliation{University of California, Los Angeles, California 90095, USA}
\author{Chan Joshi}
\email{cjoshi@ucla.edu}
\affiliation{University of California, Los Angeles, California 90095, USA}

\begin{abstract}
Intense spatial or spatiotemporal vortex pulses from the extreme ultraviolet to soft X-ray spectral windows are expected to provide new degrees of freedom for a variety of key applications since they carry longitudinal or transverse orbital angular momentum (OAM), respectively.
Plasma-based high harmonic generation driven by a near-infrared spatial or spatiotemporal optical vortex offers a promising route to such novel light sources.
However, the energy conversion efficiency from the incident vortex beam to the vortex harmonics is rather low because of the limited driving intensities available in practice.
Here, we propose and demonstrate through simulations that by adding a readily available relativistic Gaussian pump beam as a source of energy, 
the energy conversion efficiency can be increased by several orders of magnitude. 
In addition, the proposed scheme allows independent control over the frequency and OAM of the vortex harmonics.
\end{abstract}


\maketitle

A light beam that carries an orbital angular momentum (OAM) in the direction of its propagation is often referred to as a spatial optical vortex (SOV) \cite{Light_Optical_Vorticies},
exemplified by the Laguerre-Gaussian (LG) mode \cite{Allen_1992},
which has a spirally increasing/decreasing phase in the transverse plane.
During the past three decades, SOVs have been widely studied and employed in various fields \cite{PRL_Alexander, Nature_optical_manipulation, NP_manipulation, PRL_Wang, Nature_entanglement, NP_entanglement, Vieira_PRL2, Vieira_PRL, Shi_PRL_DLA, Shi_2018, PRE_Nuter, Longman_PRR, Yipeng_PRR}.
In addition to the longitudinal OAM, recent studies have demonstrated that light can also carry a transverse OAM perpendicular to its propagation direction \cite{STOV_PRA, STOV_physical_report, STOV_PRX, STOV_OPTICA, PRL_STOV, STOV_NP, PRL_STOV2}.
Unlike the conventional SOV, such transverse-OAM-carrying light has characteristic spiral phase structure in the space-time plane and is thus referred to as the spatiotemporal optical vortex (STOV).
Due to such unique features, the STOV pulses 
provide novel geometries and functionalities for vortex states, and extend their potential applications to the space-time domain.

Currently, both SOV and STOV beams are typically produced and utilized 
in the visible and infrared (IR) regimes.
Extending their frequencies to shorter wavelengths,
such as the extreme ultraviolet (EUV) and soft X-ray (SXR) spectral regions, 
will not only expand the capability of existing applications \cite{Sakdinawat_07}, 
but also open up completely new research directions in a wide range of areas.

High harmonic generation (HHG) from gases driven by a near-IR 
SOV or STOV laser beam has been shown to produce EUV and SXR pulses with longitudinal or transverse OAM \cite{NP_Zurch, PRL_Carlos, PRL_Genevieve, NC_Geneaux, STOV_gas_HHG}.
However, the SOV/STOV beams involved are limited to a relatively low intensity (typically $\lesssim 10^{14}$ W/cm$^2$) to avoid strong gas ionization, thereby resulting in a relatively low yield of vortex harmonics.
To break this limitation, 
plasma-based HHG has been proposed to generate intense 
spatial or spatiotemporal vortex harmonics.
Nevertheless, except for a few methods that employ an ultra-intense ($>10^{18}$ W/cm$^2$) near-IR Gaussian laser irradiating 
a structured overdense plasma \cite{NatPhy_Leblanc, NC_Wang, PRL_Yi, trines2023laser}, 
most proposed methods rely on the reflection of an ultra-intense ($>10^{18}$ W/cm$^2$) near-IR SOV or STOV laser off 
flat overdense plasma surfaces \cite{PRL_Vortex_HHG, PRL_Denoeud, HPL_STOV_HHG}.
Such ultrahigh-intensity vortex drivers are not widely available in practice, especially for STOV pulses.
A reduction in the intensity of the SOV/STOV driving pulse leads to a dramatic reduction of the HHG efficiency,
which in turn severely limits their applicability.
In addition, 
due to the conservation of OAM for a single driving beam,
the topological charge of the vortex harmonics is a multiple of the harmonic order and cannot be tuned independently. Therefore, this method cannot produce high harmonics with tunable low-order OAMs, 
which are preferable for many applications.

Here, we propose a novel plasma-based HHG scheme that can generate both intense spatial and spatiotemporal vortex harmonics with an arbitrary topological charge for any harmonic order. 
Our scheme combines an ultra-intense ($>10^{18}$ W/cm$^2$) Gaussian `pump' pulse and a readily available weak ($10^{15}$-$10^{17}$ W/cm$^2$) SOV or STOV `seed' pulse with orthogonal polarizations.
Since the Gaussian mode is the usual spatial mode from high-power lasers, such intense Gaussian pump beams are widely available \cite{HPL_2015, HPL_2019}.
With the addition of the Gaussian pump, 
the intensity of the vortex harmonics can be enhanced by several orders of magnitude compared to the case without the Gaussian pump.

\begin{figure}[tp]
\centering\includegraphics[height=0.41\textwidth]{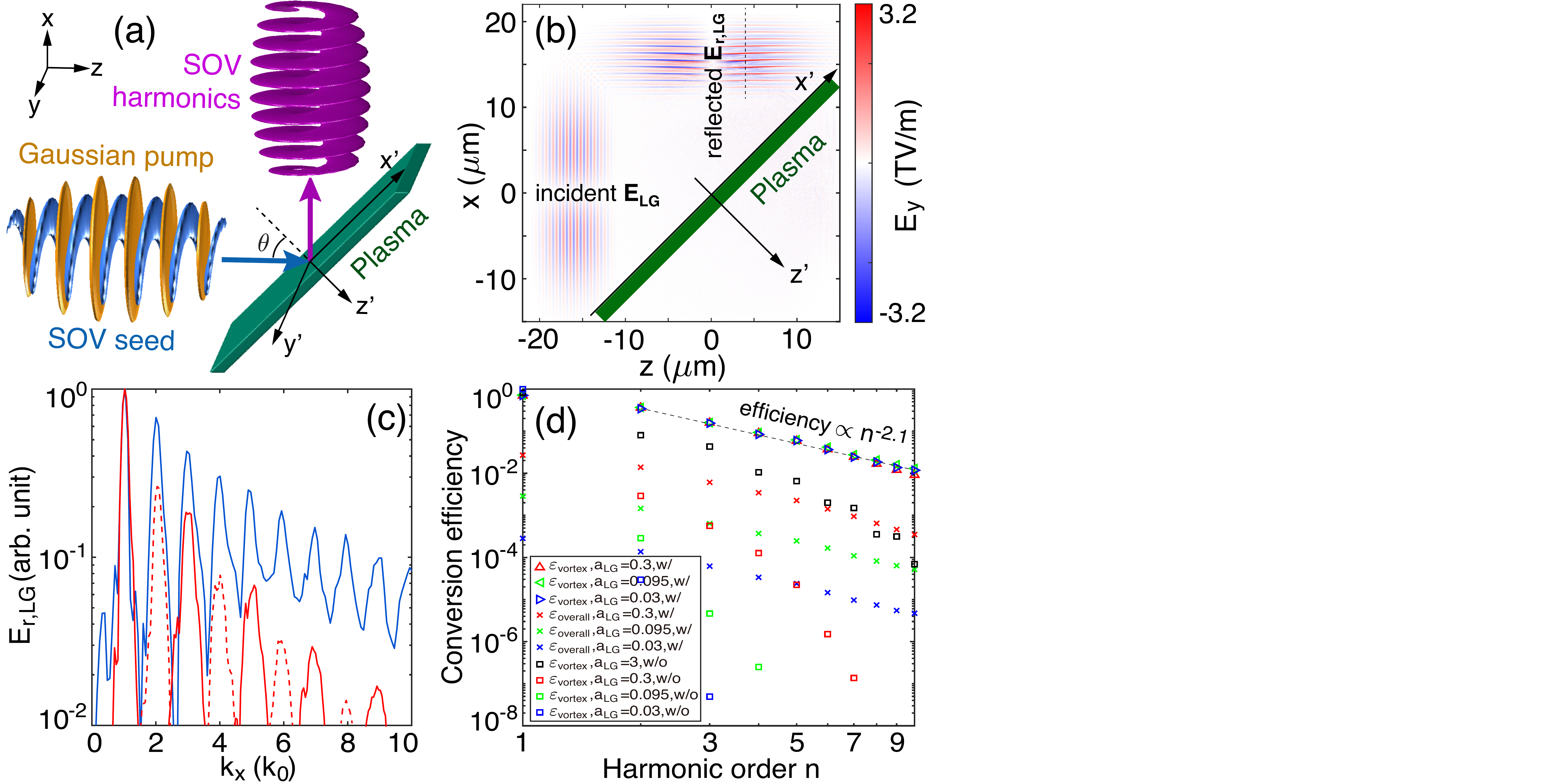}
\caption{\label{fig1} 
(a) Simulation setup.
(b) Electric fields of the incident ($\boldsymbol{E_{LG}}$) and reflected ($\boldsymbol{E_{r,LG}}$) vortex pulse observed in the $x-z$ plane at $y=0$.
(c) The blue curve shows the FFT spectrum of $\boldsymbol{E_{r,LG}}$ obtained along the black dotted line ($z=4\ \mu$m) in (b). 
For comparison, we also plot the FFT spectrum (red curve) without the Gaussian pump but with a hundredfold increase in the LG seed intensity ($a_{LG}=3$).
Note that when the Gaussian pump is absent, the odd harmonics have $s$-polarization (solid red curve) while the even harmonics have $p$-polarization (dashed red curve).
(d) Energy conversion efficiencies $\varepsilon_{vortex}$ and $\varepsilon_{overall}$ for different-order vortex harmonics with (w/) and without (w/o) the Gaussian pump.
}
\end{figure}

We first illustrate our scheme using three-dimensional (3D) particle-in-cell (PIC) simulations with the code OSIRIS \cite{Fonseca_2002}.
For simplicity, we begin with the SOV case.
The simulation setup is shown in Fig. \ref{fig1}(a).
A $p$-polarized Gaussian pump and a $s$-polarized LG SOV seed co-propagate along the $+z$ axis and then are obliquely incident on an overdense plasma slab with an incident angle of $\theta=45^\circ$,
so that the reflected pulses propagate along the $+x$ axis.
Considering the interaction occurring in a distance much less than the Rayleigh length 
and thus neglecting the wavefront curvature, 
the electric field of the Gaussian pump is expressed as
\begin{equation}\label{eq1}
\begin{aligned}
\boldsymbol{E_{G}}=\boldsymbol{e_x}E_{G}\exp(-r^2/W_G^2)\cos^2(\pi \xi/c\tau_{G})\cos(\omega_Gt-k_Gz+\psi_G)
\end{aligned}
\end{equation}
while the laser field of the LG seed (assuming zero radial mode index) is described as
\begin{equation}\label{eq2}
\begin{aligned}
\boldsymbol{E_{LG}}=&\boldsymbol{e_y}C_{LG,l} E_{LG}(\sqrt{2}r/W_{LG})^{|l_{LG}|}\exp(-r^2/W_{LG}^2)\\
&\times \cos^2(\pi \xi/c\tau_{LG})\cos(\omega_{LG}t-k_{LG}z+l_{LG}\phi+\psi_{LG})
\end{aligned}
\end{equation}
where $\boldsymbol{e_x}$ ($\boldsymbol{e_y}$) is the unit vector in the $x$ ($y$) direction, $E_{G}$ ($E_{LG}$) is the peak electric field strength of the Gaussian (LG) laser,
 $C_{LG}$ is a normalization constant,
 $l_{LG}$ is the spatial topological charge, 
 $\phi=\arctan(y/x)$ is the azimuthal angle in the transverse plane, 
and $\xi=ct-z$ is the comoving coordinate with $\xi=0$ corresponding to the pulse center. 
In the simulations,
both incident pulses have the same frequency ($\omega_G=\omega_{LG}=\omega_0=2\pi c/\lambda_0$), wave number ($k_G=k_{LG}=k_0=\omega_0/c$), pulse duration ($\tau_G=\tau_{LG}=38$ fs),
and initial phase ($\psi_{G}=\psi_{LG}=0$), with $\lambda_0=0.8\ \mu$m the laser wavelength.
The Gaussian pump has a spot size of $W_G=8\ \mu$m and a peak electric field of $E_{G}=12$ TV/m.
This corresponds to a peak normalized vector potential of $a_G=eE_G/m_ec\omega_G=3$, a peak intensity of $I_G= 2\times 10^{19}$ W/cm$^2$, and a peak power of $P_{G}=20$ TW.
The LG seed has $l_{LG}=1$, $W_{LG}=7\ \mu$m and $E_{LG}=1.2$ TV/m, corresponding to $a_{LG}=eE_{LG}/m_ec\omega_{LG}=0.3$, $I_{LG}= 2\times 10^{17}$ W/cm$^2$, and $P_{LG}=0.8$ TW \cite{Longman_LG_power}.
The plasma slab has a sharp boundary with a thickness of $2\ \mu$m and a uniform density of $4n_c$, 
where $n_c=1.74\times10^{21}cm^{-3}$ is the critical plasma density. 
The simulation box has a dimension of $38\times38\times32\ \mu m^3$ with $4750\times 4750 \times 800$ cells in the $z$, $x$ and $y$ directions, respectively. 
The plasma electrons are sampled by 2 macroparticles per cell and the ions are assumed to be immobile.

The more-intense $p$-polarized Gaussian pump predominantly drives relativistic oscillations of the plasma electron surface layer around the immobile ion background (see Supplementary Note 1 for details).
This oscillating electron surface layer then `reflects' the $s$-polarized LG seed.
Although the $p$-polarized Gaussian pump is also reflected,
it can be filtered out by using a broadband all-reflective HHG polarizer \cite{HHG_polarizer1, HHG_polarizer2} in practice. 
The phase of the reflected vortex field $\boldsymbol{E_{r,LG}}$ depends on the position of the plasma surface at the moment of the reflection. This retardation effect gives rise to a distorted waveform rich in harmonics of the fundamental frequency [Fig. \ref{fig1}(b)].
Figure \ref{fig1}(c) presents the FFT spectrum (blue curve) of $\boldsymbol{E_{r,LG}}$, which contains both odd and even harmonics. 
By spectral filtering $\boldsymbol{E_{r,LG}}$ in the frequency range [$n$-0.5, $n$+0.5]$\omega_0$, we can extract each harmonic of order $n$.
The energy conversion efficiency $\varepsilon_{vortex}$ from the vortex seed to the vortex harmonics is shown in Fig. \ref{fig1}(d) (triangles),
which can be fitted by $n^{-2.1}$.  
By introducing the incident pump to seed energy ratio, $\alpha=\frac{P_G\tau_G}{P_{LG}\tau_{LG}}$, 
we define the overall energy conversion efficiency as $\varepsilon_{overall}=\frac{1}{1+\alpha}\varepsilon_{vortex}$ and plot it in Fig. \ref{fig1}(d) (crosses).
When the seed intensity $I_{LG}$ is reduced to $2\times 10^{16}$W/cm$^2$ ($a_{LG}=0.095$) or $2\times 10^{15}$W/cm$^2$ ($a_{LG}=0.03$), 
although $\varepsilon_{overall}$ decreases accordingly, $\varepsilon_{vortex}$ remains almost unchanged.
For comparison, we also plot $\varepsilon_{vortex}$ in Fig. \ref{fig1}(d) (squares) in the absence of the Gaussian pump,
which significantly decreases as $I_{LG}$ ($a_{LG}$) decreases.
In addition, not only $\varepsilon_{vortex}$ but also $\varepsilon_{overall}$ in the two-pulse scheme are significantly higher than $\varepsilon_{vortex}$ in the single-pulse scheme with identical $I_{LG}$.
Even if $I_{LG}$ is increased 
to $2\times 10^{19}$W/cm$^2$ ($a_{LG}=3$), $\varepsilon_{vortex}$ in the single-pulse scheme is still more than two orders of magnitude lower than that in the two-pulse scheme with $I_{LG}=2\times 10^{17}$W/cm$^2$ for harmonics with $n>9$ [see the red curve in Fig. \ref{fig1}(c) and the black squares in Fig. \ref{fig1}(d)], leading to a much lower \textit{absolute} harmonic intensity in the EUV and SXR regions.

\begin{figure}[tp]
\centering\includegraphics[height=0.31\textwidth]{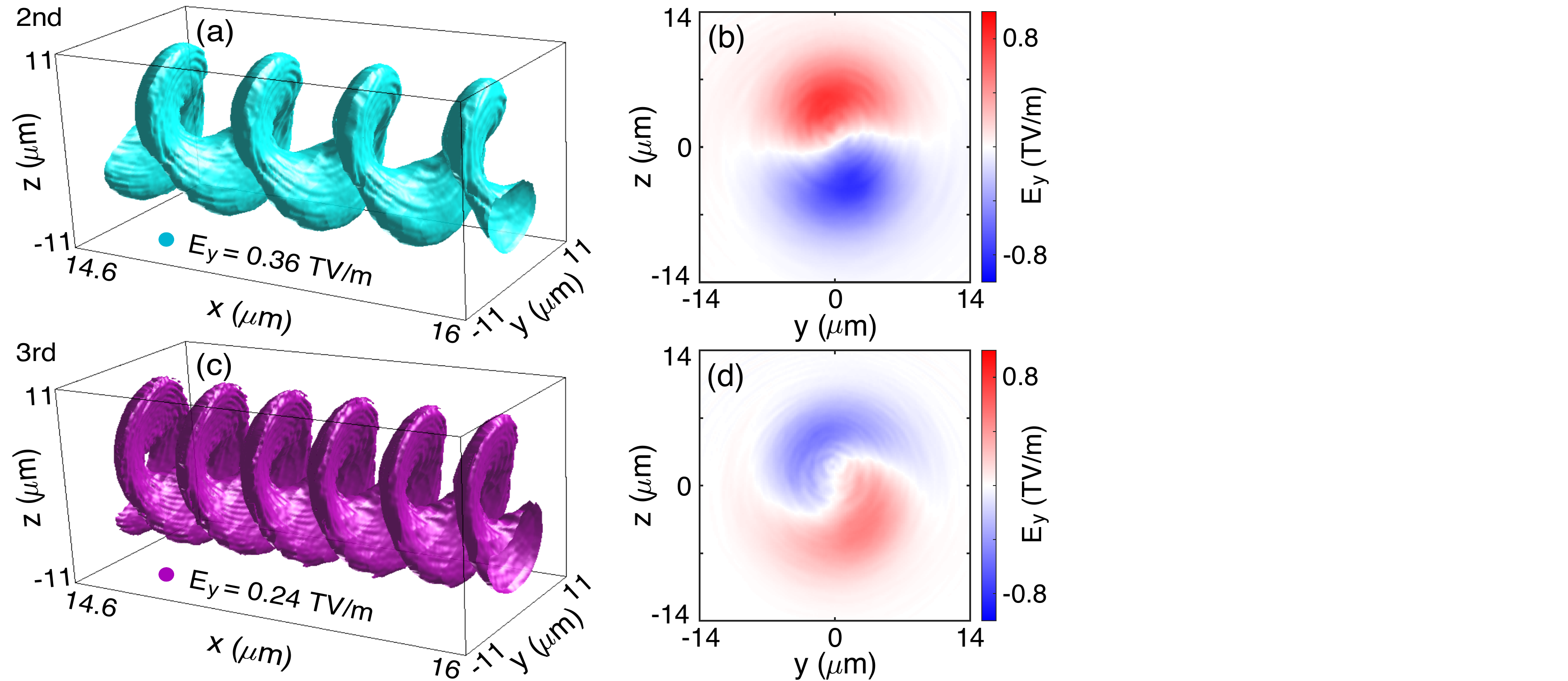}
\caption{\label{fig2} 
3D isosurfaces (left column) and 2D slices (right column) of the electric field of the 2nd (top row) and 3rd (bottom row) vortex harmonics.
}
\end{figure}

In Fig. \ref{fig2} we plot the 3D isosurfaces and transverse distributions of the 2nd and 3rd vortex harmonics.
All of them feature a single-twisted helical structure,
which means that the topological charges of these harmonics are identical to the incident value, i.e., $l_n=l_{LG}=1$.
Such a promising feature allows control over $l_n$ to an arbitrary number for any harmonic order simply by tuning $l_{LG}$ [see Supplementary Note 2 for simulation results with $l_{LG}=2$ and $3$].
This is quite different from the scheme of a single LG driving pulse where $l_n=nl_{LG}$ \cite{PRL_Vortex_HHG} and thus only high values of topological charge can be obtained for high harmonic orders.

In the following, we carry out theoretical analyses to explain the above mechanism.
To simplify the analysis,
we first rotate the $xyz$ coordinate clockwise by $\theta=45^\circ$ around the $y$ axis to the $x'y'z'$ coordinate [see Fig. \ref{fig1}(a)]
and then perform a Lorentz transformation from the laboratory frame to a new frame (the $x''y''z''$ coordinate) which moves along the $x'$ axis with a velocity of $c\sin\theta$ \cite{POF_oblique_incidence}. 
In the moving frame (quantities denoted by $''$), the incident laser propagates normally to the plasma surface, while the plasma electrons and ions move with the velocity $\boldsymbol{v''}=-c \sin\theta\boldsymbol{e_x''}$ along the $x''$-axis. 
The frequency and the wavenumber of the incident Gaussian (LG) laser change to $\omega_G''=\omega_G\cos\theta$ ($\omega_{LG}''=\omega_{LG}\cos\theta$) and $k_G''=k_G\cos\theta$ ($k_{LG}''=k_{LG}\cos\theta$), respectively.
In the $z''$ direction,
the plasma electron surface layer relativistically oscillates with the incident frequency $\omega_G''$ since the dominant driving force that electrons feel is the Lorentz $\boldsymbol{v''} \times \boldsymbol{B''}$ force due to the $p$-polarized Gaussian pump. 
This oscillating electron surface layer then acts as a mirror to reflect the $s$-polarized vortex seed.

According to the simplified oscillating mirror model \cite{IEEE_Mori, POP_Bulanov, POP_Lichters, NJP_Tsakiris, RMP_HHG},
the reflected vortex field $\boldsymbol{E_{r,LG}''}$ observed at ($t''$, $z'' < 0$) is emitted at a retarded time $t_{ret}'' = t''-Z_m''(t_{ret}'')/c+z''/c$ from the plasma mirror located at $Z_m''(t_{ret}'')$.
Therefore $\boldsymbol{E_{r,LG}''}$ at the observer is
\begin{equation}\label{eq3}
\begin{aligned}
&\boldsymbol{E_{r,LG}''}\sim \boldsymbol{e_x''}E_{LG}''\cos[\omega_{LG}''t''+k_{LG}''z''+2k_{LG}''Z_m''(t_{ret}'')\\
&+l_{LG}''\phi''+\psi_{LG}'']
\end{aligned}
\end{equation}
where $l_{LG}''=l_{LG}$, $\phi''\equiv \arctan(\frac{y''}{x''+\xi''\sin\theta})=\phi$ and $\psi_{LG}''=\psi_{LG}$.
To a first approximation, the longitudinal mirror motion can be simply expressed as
\begin{equation}\label{eq4}
\begin{aligned}
Z_m''(t_{ret}'') \approx \hat{Z_m''}(t'') \approx A_m'' \cos(\omega_G''t''+\psi_m'')
\end{aligned}
\end{equation}
where $A_m''$ and $\psi_m''$ are the amplitude and phase of the mirror oscillation, respectively. 
Substituting Eq. \eqref{eq4} into Eq. \eqref{eq3} and then using the Jacobi-Anger expansion \cite{Cuyt_2008} yield
\begin{equation}\label{eq5}
\begin{aligned}
&\boldsymbol{E_{r,LG}''} \sim  \boldsymbol{e_x''}E_{LG}'' \Re  \bigg \{ \sum_{n=1}^{\infty} i^{n-1} J_{n-1}(\eta'') \exp \Big \{  \big [(n-1)\omega_G''+\omega_{LG}'' \big ]t'' \\
&+\big [(n-1) k_G'' + k_{LG}'' \big ]z''  +  n \psi_m'' +l_{LG}''\phi''  +(\psi_{LG}''-\psi_m'') \Big \} \bigg \}
\end{aligned}
\end{equation}
where $\eta''=2k_G''A_m'' \ll 1$ and $J_{n-1}$ denotes the Bessel function of the first kind of order $n-1$.
Note that in the last step of the derivation of Eq. \eqref{eq5}, the terms which are much smaller than $J_{n-1}(\eta'')$ are neglected.
The $n$th harmonic has a frequency of $\omega_n''=(n-1)\omega_G''+\omega_{LG}''$,
and its conversion efficiency $\varepsilon_{vortex}$ depends on $J_{n-1}(\eta'')$, 
which increases with increased pump intensities (see Supplementary Note 3 for details).
All these harmonics have topological charges identical to the incident value. 
As a comparison, in the single-LG-beam case, since the oscillating plasma mirror is driven by the LG laser itself, 
$Z_m''$ is $\phi''$-dependent
and thus the azimuthal phase of the vortex harmonics changes accordingly, 
leading to a change of topological charge. 

We can also use the photon picture of light and the conservation laws of energy and OAM value to explain the above mechanism.
For most of the $n$th vortex harmonic photons in our scheme, 
according to $\omega_n''=(n-1)\omega_G''+\omega_{LG}''$ in Eq. \eqref{eq5}, 
each harmonic photon is produced from $n-1$ photons of the Gaussian pump and $1$ photon of the LG seed, thereby obtaining OAM of $l_{LG} \hbar$.

\begin{figure}[tp]
\centering\includegraphics[height=0.185\textwidth]{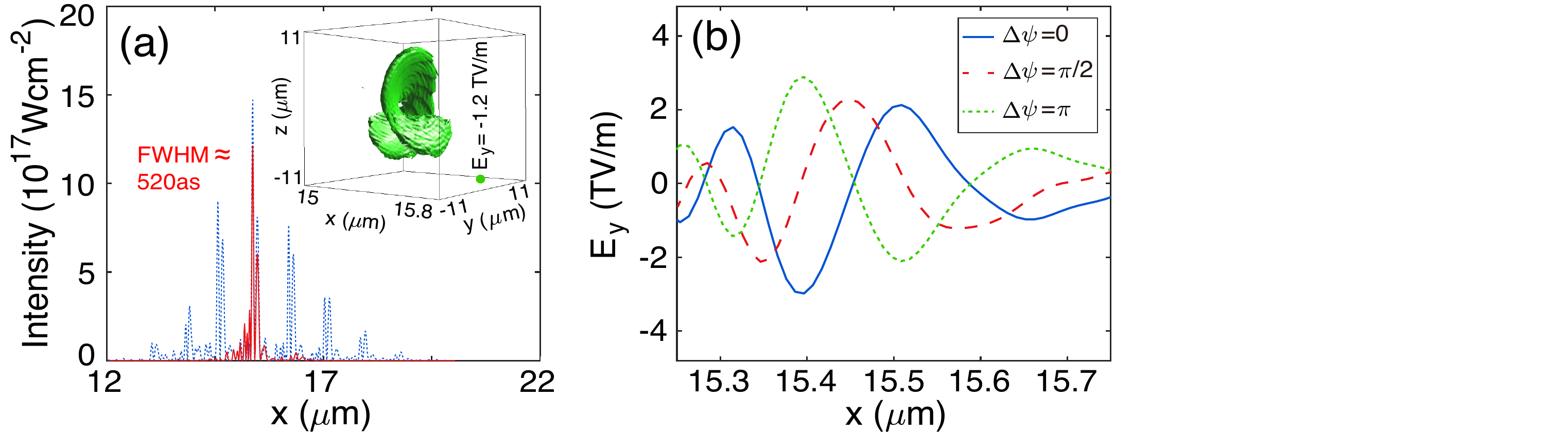}
\caption{\label{fig3} 
(a) Intensity distribution (blue curve: attosecond pulse train; red curve: single attosecond pulse) of the reflected vortex pulse filtered from 2nd to 8th harmonics observed at $y=0$ and $z=4\mu$m. The inset shows the electric field isosurface of the single attosecond pulse.
(b) Electric field lineouts of the single attosecond pulse observed at $y=0$ and $z=4\mu$m for different $\Delta \psi$ in steps of $\pi/2$.
The phases of the attosecond pulses are also shifted by approximately $\pi/2$. 
}
\end{figure}

The superposition of vortex harmonics with different orders leads to a helical attosecond pulse train in the time domain.
Figure \ref{fig3}(a) shows the intensity distribution (blue curve) of the reflected vortex pulse filtered from 2nd to 8th harmonics,
from which several attosecond pulses can be clearly seen. 
The peak intensity of these attosecond pulses is $1.5\times10^{18}$ W/cm$^2$,
about 7.5 times larger than the intensity of the incident LG seed.
The duration of the pulse train $\tau_{pulse}$ depends on the overlapping time $\tau_{overlap}$ between the Gaussian pump and the LG seed, compared with the single-pulse scheme where $\tau_{pulse}$ is essentially $\tau_{LG}$.
If we reduce $\tau_G$ to 8 fs while keep the relatively long $\tau_{LG}$ unchanged, $\tau_{overlap}$ also decreases to 8 fs and an isolated attosecond ($\sim$520 as) vortex pulse with peak intensity as high as $1.2\times10^{18}$ W/cm$^2$ can be obtained [see the 3D electric field structure (inset) and the intensity lineout (red curve) in Fig. \ref{fig3}(a)]. 
Considering its spot size is basically identical to that of the LG seed, its peak power reaches $\sim$5 TW.
Furthermore, 
its carrier-envelope-phase can be tuned by controlling the initial phase difference $\Delta \psi=\psi_{LG}-\psi_{G}$ between the Gaussian pump and the LG seed, 
as shown in Fig. 3(b).

\begin{figure}[tp]
\centering\includegraphics[height=0.35\textwidth]{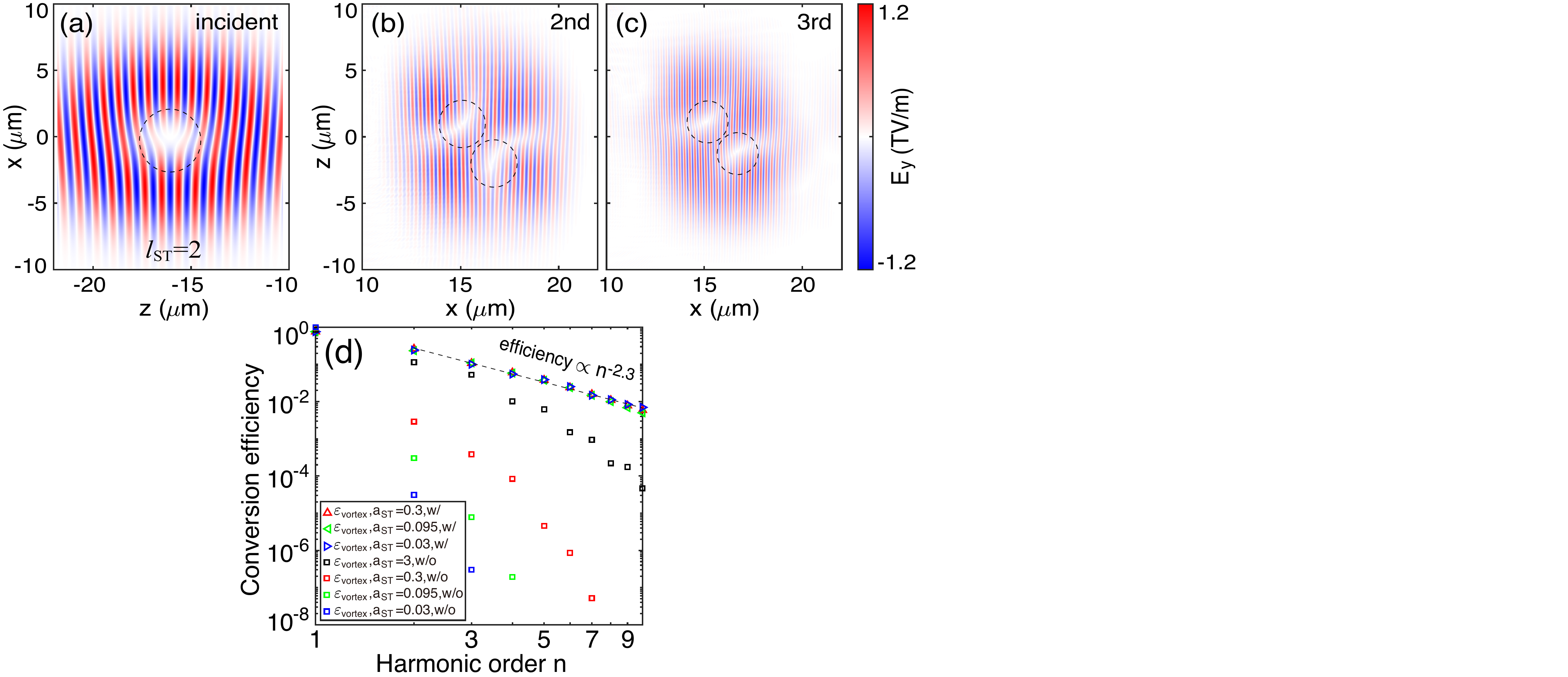}
\caption{\label{fig4} 
Electric field distributions of the incident STOV pulse (a) and the reflected 2nd (b)-3rd (c) harmonics at $y=0$.
(d) Energy conversion efficiencies $\varepsilon_{vortex}$ for different-order STOV harmonics with (w/) and without (w/o) the Gaussian pump. 
}
\end{figure}

Next, we present 3D simulation results of the proposed scheme for STOV pulses.
The parameters of the Gaussian pump and overdense plasma are identical to those used in Fig. \ref{fig1}(a),
while the seed is changed 
to a $s$-polarized STOV beam, with the electric field described as
\begin{equation}\label{eq7}
\begin{aligned}
&\boldsymbol{E_{ST}}=\boldsymbol{e_y}E_{ST}C_{ST}[(\xi/W_\xi)^2+(x/W_x)^2]^{|l_{ST}|/2}\\
&\times \exp[-(x^2/W_x^2+y^2/W_y^2+\xi^2/W_\xi^2)]\\
&\times \cos(\omega_{ST}t-k_{ST}z+l_{ST}\phi_{ST}+\psi_{ST})
\end{aligned}
\end{equation}
where $C_{ST}$ is a normalization constant,
$l_{ST}$ is the spatiotemporal topological charge, and $\phi_{ST}=\arctan(xW_{\xi}/\xi W_x)$ is the azimuthal angle in the space-time plane.
In Fig. \ref{fig4}(a), 
we plot the electric field distribution of the incident STOV pulse at $y=0$ with $\omega_{ST}=\omega_0$, $k_{ST}=k_0$, $l_{ST}=2$, $\psi_{ST}=0$, and $W_{\xi}=W_x=W_y=4\mu$m. 
This STOV pulse has a peak electric field of $E_{ST}=1.2$ TV/m, corresponding to $a_{ST}=eE_{ST}/m_ec\omega_{ST}=0.3$ and $I_{ST}= 2\times 10^{17}$W/cm$^2$. 
The lower part ($x<0$) of its electric field has two less periods than the upper part ($x>0$) and this results in a relative up-and-down phase difference of $4\pi$ (corresponding to $l_{ST}=2$), forming a fork-grating-shaped field structure [see Fig. \ref{fig4}(a)].

Similar to the LG seed case, 
$s$-polarized spatiotemporal vortex harmonics can be generated when the incident $s$-polarized STOV seed is reflected from the oscillating plasma surface.
Figures \ref{fig4}(b)-\ref{fig4}(c) present the 2nd and 3rd harmonics, 
where two fork dislocations (within the dotted circle) for both harmonics can be observed.
Each fork dislocation shows an up-and-down phase difference of $2\pi$, leading to a total spatiotemporal topological charge of $l_n=l_{ST}=2$.
As a comparison, in the case of a single STOV driving beam, the spatiotemporal topological charge is $n\times l_{ST}$ for the $n$th harmonic \cite{HPL_STOV_HHG}.

The two-pulse oscillating mirror model in the moving frame can also be used to explain the underlying physics of STOV harmonic generation.
By replacing $\phi''$ with $\phi_{ST}''$ in Eq. \eqref{eq5} (see Supplementary Note 4 for details),
we can obtain the reflected STOV field:
\begin{equation}\label{eq8}
\begin{aligned}
&\boldsymbol{E_{r,ST}''} \sim  \boldsymbol{e_x''}E_{ST}'' \Re  \bigg \{ \sum_{n=1}^{\infty} i^{n-1} J_{n-1}(\eta'') \exp \Big \{  \big [(n-1)\omega_G''+\omega_{ST}'' \big ]t'' \\
&+\big [(n-1) k_G'' + k_{ST}'' \big ]z''  +  n \psi_m'' +l_{ST}''\phi_{ST}''  +(\psi_{ST}''-\psi_m'') \Big \} \bigg \}
\end{aligned}
\end{equation}
where $\omega_{ST}''=\omega_{ST}\cos\theta$, $k_{ST}''=k_{ST}\cos\theta$, $l_{ST}''=l_{ST}$, and $\psi_{ST}''=\psi_{ST}$.
Equation \eqref{eq8} shows that all the harmonics have the same spatiotemporal topological charge as the incident value. 
Alternatively, from the viewpoint of simultaneous conservation of photon energy and OAM value,
each photon of the $n$th STOV harmonic is generated from $n-1$ photons of the Gaussian pump and 1 photon of the STOV seed, thereby finally carrying transverse OAM of $l_{ST}\hbar$.

The energy conversion efficiencies $\varepsilon_{vortex}$ from the STOV seed to different-order vortex harmonics with $I_{ST}$ varied from $2\times 10^{17}$W/cm$^2$ ($a_{ST}=0.3$) to $2\times 10^{15}$W/cm$^2$ ($a_{ST}=0.03$) are shown in Fig. \ref{fig4}(d) (triangles). 
As one can see, 
$\varepsilon_{vortex}$
is almost independent of $I_{ST}$ within the above intensity range.
The power-law scaling can be fitted by $n^{-2.3}$.  
As a comparison, when the Gaussian pump is absent, 
$\varepsilon_{vortex}$ [see squares in Fig. \ref{fig4}(d)]
decreases dramatically with decreasing $I_{ST}$, and it is 
significantly lower than that in our proposed scheme with the Gaussian pump.
In addition,
even though $I_{ST}$ in the single-pulse scheme is increased to $2\times 10^{19}$W/cm$^2$ ($a_{ST}=3$),
both the conversion efficiency and the \textit{absolute} intensity value remain lower than those in the two-pulse scheme with $I_{ST}=2\times 10^{17}$W/cm$^2$ for harmonics ($n>9$) in the EUV and SXR ranges.

In summary,
we have proposed a novel two-pulse pump-seed scheme that can efficiently generate both spatial and spatiotemporal vortex harmonics at short wavelengths. 
Our scheme also enables independent control over the OAM and frequency of the vortex harmonics.
Such novel light sources can open up a variety of key applications.




\section{references}


\end{document}